\documentclass[reprint,aps,prl,nofootinbib,superscriptaddress]{revtex4-2}
\usepackage[utf8]{inputenc}
\usepackage{amsmath}
\usepackage{amsfonts}
\usepackage{amsthm}
\usepackage{amssymb}
\usepackage{graphicx}
\usepackage[usenames,dvipsnames]{color}
\usepackage{hyperref}
\usepackage{mathrsfs}%cursive font
\usepackage{orcidlink}

\theoremstyle{definition}

\theoremstyle{remark}

\newcommand{\norm}[1]{\left\Vert#1\right\Vert}
\newcommand{\abs}[1]{\left\vert#1\right\vert}
\newcommand{\set}[1]{\left\{#1\right\}}

\newcommand{\eps}{\varepsilon}

\newcommand{\expec}[1]{\left\langle#1\right\rangle}

\newcommand{\zhat}{\mathbf{\hat{z}}}
\newcommand{\yhat}{\mathbf{\hat{y}}}
\newcommand{\xhat}{\mathbf{\hat{x}}}
\newcommand{\ket}[1]{\left|#1\right\rangle}
\newcommand{\bra}[1]{\left\langle#1\right|}
\newcommand{\kpsi}{\ket{\psi}}
\newcommand{\bpsi}{\bra{\psi}}

\def\actaa{\ref@jnl{Acta Astron.}}      % Acta Astronomica
\usepackage[normalem]{ulem}

\newcommand{\chg}[1]{\textcolor{black}{#1}}

\begin{document}

\title{Constraining Axion Dark Matter with Galactic-Centre Resonant Dynamics}

\author{Yonadav Barry Ginat\,\orcidlink{0000-0003-1992-1910}}%
\email{yb.ginat@physics.ox.ac.uk}%
\affiliation{Rudolf Peierls Centre for Theoretical Physics, University of Oxford, Parks Road, Oxford, OX1 3PU, United Kingdom}%
\affiliation{New College, Holywell Street, Oxford, OX1 3BN, United Kingdom}%
\author{Bence Kocsis}
\affiliation{Rudolf Peierls Centre for Theoretical Physics, University of Oxford, Parks Road, Oxford, OX1 3PU, United Kingdom}
\affiliation{St.~Hugh's College, St.~Margaret's Road, Oxford, OX2 6LE, United Kingdom}


\begin{abstract}
We study the influence of \chg{fuzzy-}dark-matter cores on the orbits of stars at the Galactic centre. This dark matter candidate condenses into dense, solitonic cores, and, if a super-massive black hole is present at the centre of such a core, its central part forms a `gravitational atom'. Here, we calculate the atom's contribution to the gravitational potential felt by a Galactic-centre star, for a \chg{general} state of the atom. We study the angular-momentum dynamics this potential induces, and show that it is similar to vector resonant relaxation. Its influence is found to be \chg{potentially} sufficiently strong that such a dynamical component should be accounted for in Galactic-centre modelling. For the Milky Way, the atom is expected to have some spherical asymmetry, and we use this to derive a stability condition for the disc of young, massive stars at the Galactic centre---if the atom's mass is too large, then the disc would be destroyed. Thus, the existence of this disc constrains the mass of the particles comprising the solitonic core\chg{. We study an example model of the core, where all of the rotation of the core's inner region is assumed to come from an $l=1$ state, and its amplitude is determined by the halo's spin parameter;} such a core is found to be in tension with the \chg{stability} of the clockwise stellar disc for $4.\chg{2}\times 10^{-20}\,\textrm{eV} \leq m_a \leq 5.\chg{4}\times 10^{-20}\,\textrm{eV}$ at $2\sigma$. \chg{Other core models would vary the constrained values of $m_a$ somewhat.} These constraints will tighten significantly with future, improved data. 
\end{abstract}

\maketitle

\paragraph{Introduction.}
Fuzzy dark matter (FDM), made of axions (or axion-like particles), is a leading candidate for a dark-matter constituent particle \cite{AbbottSikivie1983,DineFischler1983,Preskilletal1983,Huietal2017,BarOr+2019,Hui2021,BarOr+2021,Marshetal2024}. As these particles are bosons, the dark-matter haloes they form are Bose--Einstein condensates \cite{Schiveetal2014a,Schiveetal2014b,Marshetal2024}, and the axions all inhabit a single \chg{profile}, with an extremely large occupation number, over astrophysical scales. An FDM halo, therefore, is characterised by a single \chg{field} $\psi$ (normalised by $\int \abs{\psi}^2 \mathrm{d}^3 x = 1$), which solves the Schr\"{o}dinger--Poisson system in the non-relativistic limit \cite[e.g.][]{Huietal2017,Marshetal2024}.
%\begin{align}
%    \label{eqn:Schroedinger}\mathrm{i}\hbar\frac{\partial \psi}{\partial t} & = -\frac{\hbar^2}{2m_a}\nabla^2\psi + V\psi, \\
%    \label{eqn:Poisson} \nabla^2 V & = 4\pi G M_{\rm vir} \abs{\psi}^2,
%\end{align}
%where $m_a$ is the particle mass and $M_{\rm vir}$ is the total mass of the system ($\psi$ is normalised so that $\int \abs{\psi}^2 \mathrm{d}^3 x = 1$). 

%We consider individual particle masses $m_a$ in the range $10^{-20}\, \mathrm{eV} \lesssim m_a \lesssim 10^{-1\chg{9}} \, \mathrm{eV}$, which correspond to de-Broglie wave-lengths of the order of a parsec \chg{(with speeds of order of the orbital speed at the bottom of the potential well)}. This range lies above the mass range constrained by the Ly$\alpha$ forest \cite[\chg{$m_a \lesssim 2\times 10^{-20}\, \textrm{eV}$;}][]{RogersPeieris2021} \chg{or dwarf galaxies \cite[$m_a \lesssim 2\times 10^{-21}\, \textrm{eV}$;][]{Zimmermannetal2024}}, but below the masses best constrained by the precession of the orbit of S2 \cite[\chg{$10^{-19}\, \textrm{eV} \lesssim m_a \lesssim 10^{-18}\, \textrm{eV}$}][]{GRAVITY2019,GRAVITY2023} and is largely unconstrained. 
In this paper, we study potential constraints on the existence of FDM, from \chg{the long-term} dynamics of stars at the Galactic centre, in particular from the dynamics of angular momenta of young, massive stars that lie in a disc around the super-massive black hole (SMBH) there 
\cite{LevinBeloborodov2003,PaumardEtAl2006,Bartkoetal2009,Yelda2014,Schoedeletal2020,Fellenbergetal2022}\chg{, }
%\cite{LevinBeloborodov2003,PaumardEtAl2006,Bartkoetal2009,Yelda2014,Gallego-Canoetal2018,Alietal2020,Schoedeletal2020,Fellenbergetal2022};
\chg{which are governed by vector resonant relaxation (VRR) \cite{RauchTremaine1996,KocsisTremaine2011,KocsisTremaine2015,Roupas+2017,BarOr+2018,Fouvryetal2019,Szolgyenetal2021,Fouvry+2022,Panamarev2022,Ginatetal2023,Wang_Kocsis2023,FloresFouvry2024}. This process is driven by the orbit-averaged torques between stars that execute rapid, eccentric orbits around the SMBH and undergo fast in-plane apsidal precession (due to the spherical component of the mass distribution) on time-scales of $10$--$10^3$ and $\sim10^4$ years, respectively. 
VRR governs the distribution of stellar orbital directions at the Galactic centre (e.g.~the clockwise disc and the counter-clockwise structure \cite{Genzel+2010}). To our knowledge, VRR has never been studied in the context of ultra-light dark matter; but here we demonstrate that this may probe FDM} individual particle masses $m_a$ \chg{in a range that is currently mostly unconstrained,} $10^{-20}\, \mathrm{eV} \lesssim m_a \lesssim 10^{-1\chg{9}} \, \mathrm{eV}$, \chg{corresponding} to de Broglie wave-lengths of the order of a parsec (with speeds of order of the orbital speed at the bottom of the potential well). This range lies above the mass range constrained by the Ly$\alpha$ forest \cite[\chg{$m_a \lesssim 2\times 10^{-20}\, \textrm{eV}$;}][]{RogersPeieris2021} \chg{or dwarf galaxies \cite[$m_a \lesssim 2\times 10^{-21}\, \textrm{eV}$;][]{ElZantetal2020,Zimmermannetal2024}}, but below the masses best constrained by the precession of the orbit of S2 \cite[\chg{$10^{-19}\, \textrm{eV} \lesssim m_a \lesssim 10^{-18}\, \textrm{eV}$;}][]{GRAVITY2019,Yuanetal2022,GRAVITY2023,DellaMonica_deMartino2023a,DellaMonica_deMartino2023b}.\footnote{\chg{Other works explored the effects of dynamical friction on the relaxation of the SMBH \cite[e.g.][]{Huietal2017,Bertone2024,Glennonetal2024}; dynamical friction affects the evolution of the angular-momentum vectors much less than VRR, which is a hundredfold more efficient \cite{Kocsis+2011,Szolgyenetal2021}.}}  

\chg{Interestingly, VRR has an effective Hamiltonian similar to that of liquid crystals, because the stars cover axisymmetric annuli as they orbit the SMBH, whose orbit-averaged Newtonian gravitational effect is similar to the interaction between axisymmetric liquid-crystal molecules. The observed distribution of angular-momentum vectors of stars in the innermost $0.5$ pc of the Galactic centre exhibits coherent and incoherent structures \cite{Bartkoetal2009,Yelda2014,Fellenbergetal2022}, possibly related to nematic-isotropic phase-transitions \cite{Roupas+2017,Takacs_Kocsis2018}.}

\chg{We examine how fuzzy dark matter affects this picture. We focus on stars that} orbit the SMBH in a disc of radius $\sim 0.1$ pc---of the same order as the de Broglie wave-length of the \chg{tentative} axions---and whose mass is estimated to be a few thousand solar masses \cite{Bartko2010,Yelda2014}. We show below that a fuzzy dark-matter halo leads to the formation of a rotating dynamical component of axions of comparable mass around the SMBH, within the inner parsec, and that these axions exert torques on stars at the Galactic nucleus\chg{;} that is, this extra rotating component, if indeed dark matter is primarily fuzzy, plays an important role in the dynamics of the Galactic centre, and has potential observational consequences. \chg{We derive a general framework for computing these torques, and for gauging their influence on the stability of dynamical structures. As an} example we show that for certain values of $m_a$ \chg{and certain FDM field-configurations, they} could even destroy the young stellar disc---and the disc's existence could thus be used to place constraints on FDM. These constraints complement existing ones \cite[e.g.][]{Marsh:2018zyw,Schive:2019rrw,DesjacquesNusser2019,GRAVITY2019,Rozneretal2020,Desjacquesetal2020,RogersPeieris2021,Chiang:2021uvt,GRAVITY2023,PantigOvgun2023} \chg{\cite[see][for reviews]{Marsh2016,Huietal2017,Hui2021,Marshetal2024}} and are expected to improve as Galactic centre observations improve.

\paragraph{Fuzzy-dark-matter properties.}
Haloes of fuzzy dark matter are known to form dense soliton cores at their centres \citep{Chavanis2011,Schiveetal2014b,Huietal2017,Hui2021}, where the density $\rho \propto \abs{\psi}^2$ is uniform, whose masses $M_{\rm c}$ and radii $R_{\rm c}$ are correlated with the entire halo's virial mass $M_{\rm vir}$, \emph{viz}.,
\begin{align}
	\label{eqn:core radius} R_{\rm c} & = 100 \; \textrm{pc} \times \left(\frac{10^9\; M_{\odot}}{M_{\rm c}}\right)\left(\frac{10^{-22}\; \textrm{eV}}{m_a}\right)^2, \\ 
    \label{eqn: core mass} M_{\rm c} & = 6.7\times 10^7\; M_{\odot} \times \left(\frac{10^{-22}\; \textrm{eV}}{m_a}\right)\left(\frac{M_{\rm vir}}{10^{10}\; M_{\odot}}\right)^{1/3}.
\end{align}

An axion core would serve as a source of dark matter at the centre of the halo. Thus, around the SMBH, such a concentration of axions would simply form a `gravitational atom' \cite[e.g.][]{Chavanis2019,DavisMocz2020}, which, in the non-relativistic limit, has the same eigenstates as the hydrogen atom \citep{Baumannetal2019}. This gravitational atom is parameterised by two quantities: the total atom mass, $M_{\rm a}$, and the Bohr radius
\begin{equation}\label{eqn: Bohr radius definition}
	r_{\rm B} \equiv \frac{\hbar^2}{GM_\bullet m_a^2}.
\end{equation}

Suppose that $r_{\rm B}$ is of the order of a parsec, so that it encompasses the orbits of stars in the Galaxy's inner region, and consider a star of mass $m_s$ inside the SMBH's gravitational sphere of influence. Then, the SMBH dominates the gravitational field, and the star's orbit is well-approximated by a Keplerian ellipse. Further assume that 
\begin{equation}\label{eqn:mass hierarchy}
m_s \ll M_{\rm a} \ll M_\bullet.
\end{equation}
where $M_\bullet \approx 4\times 10^6 ~M_\odot$ is the SMBH mass \cite{Schoedeletal2002,Eisenhaueretal2005,Ghezetal2005,Ghezetal2008,GravityCollaboration2018,EventHorizonTelescope2022}. Then, to leading order, the force the atom would exert on a star (a classical object) is generated by a potential 
\begin{equation}\label{eqn:potential definition}
\Phi(\mathbf{x}) = -\int \mathrm{d}^3y \frac{G\rho_{\rm a}(\mathbf{y})}{\abs{\mathbf{x}-\mathbf{y}}},
\end{equation}
where $\rho_{\rm a}(\mathbf{y}) \equiv M_{\rm a} \abs{\psi(\mathbf{y})}^2$, with $\psi$ being the \chg{field configuration} of the gravitational atom. 
Generically\chg{,} $\psi$ is a linear combination of hydrogen-atom eigenstates $\ket{nlm}$ for the usual quantum numbers $n,l,m$, 
\begin{equation}\label{eqn:wave-function}
  \kpsi = \sum_{n,l,m}\alpha_{nlm}\ket{nlm}.
\end{equation}

\paragraph{Orbit averaging.}
The hierarchy \eqref{eqn:mass hierarchy} implies that the dynamics are, to leading order, decoupled Kepler problems (both for the axions and the star); their interaction occurs on longer time-scales than the orbital times, and therefore one may orbit-average their interaction Hamiltonian \eqref{eqn:potential definition}. The interaction of orbit-averaged Keplerian ellipses is governed by resonant relaxation \cite{RauchTremaine1996,KocsisTremaine2011,KocsisTremaine2015,BarOr+2018,Panamarev2018,Fouvry+2022,Panamarev2022}, so we require the orbit-averaged value of $\Phi$, over a Keplerian orbit of the star around the SMBH. This orbit, described by the co-ordinate $\mathbf{x}$, is assumed to have orbital parameters $a,e,i,\Omega,\omega$, denoting the semi-major axis, eccentricity, inclination and the arguments of the ascending node and the pericentre, respectively. The orbit-dependent part of equation \eqref{eqn:potential definition} is $\abs{\mathbf{x}-\mathbf{y}}\chg{^{-1}}$ (where $\mathbf{y}$ is itself an integration variable); we decompose it with spherical harmonics as
\begin{equation}\label{eqn:spherical harmonics 1/r}
	\frac{1}{\abs{\mathbf{x}-\mathbf{y}}} = \sum_{l,m} \frac{4\pi}{2l+1}\frac{\min\set{x,y}^{l}}{\max\set{x,y}^{l+1}} Y_{lm}^*(\xhat)Y_{lm}(\yhat),
\end{equation}
where $x = \abs{\mathbf{x}}$, $\xhat = \mathbf{x}/x$, \emph{etc}. We time-average over all effects that take place on time-scales much shorter than those of orbital energy change, eccentricity change, and the change of orbital orientation driven by \chg{VRR}, which takes place on megayear time-scales \cite{KocsisTremaine2011,KocsisTremaine2015}. In particular, the mean anomaly and the argument of pericentre change rapidly, and one may average over them. 
%We are now faced with two possibilities: 
%\begin{enumerate}
%	\item Consider vector resonant relaxation, and average over the mean anomaly and the argument of pericentre. This will only allow for an estimate of the mass precession, but will allow for analytic expressions, following \cite{KocsisTremaine2015}. 
%	\item Not average over the argument of pericentre. This does not admit simple analytic expressions. In that case, we can view $\Phi$ as a super-position of three-body Hamiltonians, where the star is a test particle, and $\mathbf{y}$, as an integration variable, describes a circular orbit with radius $y = \abs{\mathbf{y}}$. Single-averaged Hamiltonians exist in this limit \citep{LaskarBoue2010}, but are quite involved, with simple expressions available only for the low multipoles. Furthermore, it is unclear if the three-body series converges for $a(1-e) \leq y \leq a(1+e)$.
%\end{enumerate}
%Let us proceed with option (1) at the moment. 
We calculate the double-averaged potential $\expec{\Phi}_{\rm da}$, over \chg{of these angles}, in the appendix. The result is 
\begin{equation}\label{eqn:axion potential VRR}
	\expec{\Phi}_{\rm da} = -\sum_{l,m}J_{lm}Y_{lm}^*(\chg{\hat{\mathbf{L}}}),
\end{equation}
where the coefficients $J_{lm}$ are defined in the appendix (equation \eqref{eqn:J lm}), \chg{and the chosen $\zhat$-axis for the multipole decomposition is} $\hat{\mathbf{L}}_{\rm a}$\chg{,} the \chg{direction of the atom's} angular momentum\chg{,} and $\mathbf{L}$ is the star's angular momentum.  

Equation \eqref{eqn:J lm} implies, \emph{inter alia}, that, if $\kpsi= \sum \alpha_{n_0l_0m_0}\ket{n_0l_0m_0}$ is a super-position of non-degenerate eigenstates (i.e. contains only at most one eigenstate with each $n_0$), then $J_{lm} = 0$ unless $m=0$.  
Indeed, for this state the behaviour of this potential is quite similar to that of the classical case, where one may write 
\begin{equation}\label{eqn:potential eigenvector}
	\left\langle \Phi \right\rangle_{\rm da} = - \sum_{l_0}\sum_{l=0}^{2l_0} J_{l,l_0}P_{l}\left(\hat{\mathbf{L}}\cdot \hat{\mathbf{L}}_{\rm a}\right),
\end{equation}
where $\sqrt{4\pi/(2l+1)}J_{l0} = \sum_{l_0} J_{l,l_0}$, with the contribution of the mode $\ket{n_0l_0m_0}$ denoted by $J_{l,l_0} \propto |\alpha_{n_0l_0m_0}|^2$. 
This potential induces precession about $\hat{\mathbf{L}}_{\rm a}$, given by
\begin{equation}\label{eqn:Omega eigenvector}
\begin{aligned}
  \boldsymbol{\Omega} & = -m_s\sum_{l_0}\sum_{l=0}^{2l_0} \frac{J_{l,l_0}}{L}P_l'\left(\hat{\mathbf{L}}\cdot \hat{\mathbf{L}}_{\rm a}\right)\hat{\mathbf{L}}_{\rm a} \\ & 
  = -\frac{GM_{\rm a}m_s}{L r_{\rm B}}\sum_{l_0,l} I_{l,l_0}P_l'\left(\hat{\mathbf{L}}\cdot \hat{\mathbf{L}}_{\rm a}\right)\hat{\mathbf{L}}_{\rm a},
\end{aligned}
\end{equation}
\chg{and in general
\begin{equation}\label{eqn:Omega general lm}
\begin{aligned}
  \boldsymbol{\Omega} = -\frac{m_s}{L}\sum_{lm} J_{lm}\frac{\partial Y_{lm}^*(\hat{\mathbf{L}})}{\partial \hat{\mathbf{L}}} = -\frac{GM_{\rm a}m_s}{Lr_{\rm B}}\sum_{lm} I_{lm}\frac{\partial Y_{lm}^*}{\partial \hat{\mathbf{L}}},
\end{aligned}
\end{equation}}
%Equation \eqref{eqn:Omega eigenvector} generalises to 
%\begin{equation}\label{eqn:Omega sum of non-degenerate eigenvector}
%	\boldsymbol{\Omega} = -m_s\sum_{l,l_0} \frac{J_{l,l_0}}{L}P_l'\left(\hat{\mathbf{L}}\cdot \hat{\mathbf{L}}_{\rm a}\right)\hat{\mathbf{L}}_{\rm a},
%\end{equation}
%for $\kpsi= \sum \alpha_{n_0l_0m_0}\ket{n_0,l_0,m_0}$, where $\alpha_{n_0l_0m_0}$ is such that if both $\alpha_{n_0l_0m_0}, \alpha_{n_1l_1m_1} \neq 0$, then either $n_1\neq n_0$ or all $\ket{n_0,l_0,m_0} = \ket{n_1,l_1,m_1}$. 
where we have defined the dimensionless coefficient\chg{s} %which in the approximations made below turns out to be constant
\begin{equation}\label{eqn: I definition from J}
\begin{aligned}
    & I_{l,l_0}\left(\frac{a}{r_{\rm B}},e\right)  \equiv \frac{r_{\rm B}}{GM_{\rm a}} J_{l,l_0}, \\ &
    \chg{I_{lm}\left(\frac{a}{r_{\rm B}},e\right)  \equiv \frac{r_{\rm B}}{GM_{\rm a}} J_{lm}}.
\end{aligned}
\end{equation}
Equation \chg{\eqref{eqn:Omega general lm}} implies that an axion core with some angular momentum would participate in the VRR dynamics of the Galactic centre, and thus must be accounted for in the latter's modelling. Let us exemplify this by calculating one aspect of the axions' influence on the nuclear stellar disc; to do so, we list some assumptions and estimates of the relevant parameters below\chg{, starting with the clockwise disc}.

\paragraph{Resonant disc-breaking.}
The distribution of young, massive stars at the Galactic centre consists of an inner disc, rotating clockwise, whose mass is $3000\; M_\odot \lesssim M_{\rm d} \lesssim 10^4\; M_\odot$, up to a radius of $\sim 0.4 \; $pc, and an outer, tilted disc, 
%whose mass is $5000 \; M_\odot \lesssim M_{\rm out} \lesssim 20000\; M_\odot$,  
whose inner radius is $\sim 0.4 \; $pc \cite{Bartkoetal2009,Bartko2010,Yelda2014,Fellenbergetal2022}. %\byg{citations. These are Sebastiano's estimates, but are as yet unpublished. I asked him and he told me it would be a while until he publishes. So I can either use these, and cite it as a private communication, or use weaker estimates from \cite{Bartko2010,Yelda2014}, but then obviously get more uncertainty on the axion constraints. Bence, what do you think?} \bk{not sure, private communication sounds like it is preliminary which might change. How bad is it if we use the published results in \cite{Fellenbergetal2022}}\byg{They don't have mass estimates at all in \cite{Fellenbergetal2022}, as far as I could tell.}\byg{Well, I've decided to use $2000 \leq M_{\rm d} \leq 8000$, and be conservative. Then I can cite \cite{Bartko2010,Yelda2014}, and this makes the uncertainty in the constraint larger, but also makes clear that as soon as we get more accurate disc masses, we'll have a tighter constraint.}\bk{sounds good}

%\subsection{Constraints From Disc Stability}
Having computed the potential that the gravitational atom induces, we can use it to calculate the effect that it would have on a disc of stars. Equation \eqref{eqn:Omega eigenvector} applies to every individual star in the disc, but they also feel the torques from all the other stars. Thus, whether a given star remains bound to the disc, is determined by the relative strength of the torque exerted on it by the rest of the disc, versus the gravitational atom's torque\chg{-difference within the disc}. Recently, ref.~\cite{PanamarevKocsis2024} derived a stability criterion for a disc under the influence of a potential like \eqref{eqn:potential eigenvector} by considering the tidal torque (in angular-momentum space) due to the external potential, and comparing it with the torque from the rest of the disc: if the disc, due to its thickness, has width $\Delta \mathbf{L}$ in angular-momentum space, about \chg{the total} $\mathbf{L}_{\rm d}$, then it is stable if 
\begin{equation}\label{eqn:stability condition general}
    \abs{\chg{\Big(}\Delta L\frac{\partial \boldsymbol{\Omega}}{\partial L} \times \hat{\mathbf{L}} + \frac{1}{L}\boldsymbol{\Omega} \times \Delta \mathbf{L}\chg{\Big)\cdot\Delta \hat{\mathbf{L}}}} < \abs{\boldsymbol{\Omega}_{\rm d} \times \hat{\mathbf{L}}},
\end{equation}
where $\boldsymbol{\Omega}_{\rm d}$ is the frequency of precession of the angular momenta of the disc stars, about $\mathbf{L}_{\rm d}$\chg{, and $\Delta \hat{\mathbf{L}} \equiv \Delta \mathbf{L}/\Delta L$}. The disc's thickness $\Delta L/L$ is related to its maximum opening half-angle $\iota$ as $\Delta L/L \sim 2\sin \iota$, and satisfies $\iota < \theta$\chg{; its} precession frequency is \cite{KocsisTremaine2015,Panamarev2022}
\begin{equation}\label{eqn:disc precession frequency}
  \boldsymbol{\Omega}_{\rm d} = -\sum_{l=2}^{\infty}\sum_{i \in \textrm{disc}}  \frac{\mathcal{J}_{isl}}{L}P_l'\left(\hat{\mathbf{L}}\cdot \hat{\mathbf{L}}_i\right)\hat{\mathbf{L}}_i,
\end{equation}
where $\mathcal{J}_{isl}$ are defined in \cite{KocsisTremaine2015}. 
%For a thin disc, we approximate $\boldsymbol{\Omega}_{\rm d}$ as 
%\begin{equation}
%  \boldsymbol{\Omega}_{\rm d} \approx - \sum_{l=2}^{\infty}\sum_{i \in \textrm{disc}} \frac{\mathcal{J}_{isl}}{L}P_l'(1)\hat{\mathbf{L}}_{\rm d}.
%\end{equation}
Parameterising the right-hand side of inequality \eqref{eqn:stability condition general} without loss of generality as $[Gm_sM_{\rm d}/(La_{\rm d})] \xi_{\rm d} \sin \iota$, where\chg{,} roughly, $\xi_{\rm d}$ is of order unity, one needs to have, for stability,
\chg{
%\begin{equation}\label{eqn:stability criterion completely general}
%    \norm{\sum_{lm} \frac{J_{lm} \eps_{ij}^{~~k}}{GM_{\rm d}}\left(\frac{\partial Y_{lm}^*}{\partial \hat{L}_j} \Delta L_k + \frac{\partial^2 Y_{lm}^*}{\partial \hat{L}_j\partial\hat{L}_n}\hat{L}_k\Delta L_n\right)} \leq \frac{\xi_{\rm d}}{a_{\rm d}},
%\end{equation}
%\begin{equation}\label{eqn:stability criterion completely general}
%    \norm{\sum_{lm} \frac{J_{lm}}{GM_{\rm d}}\left(\frac{\partial Y_{lm}^*}{\partial \hat{\mathbf{L}}} \times \Delta \mathbf{L} + \Delta L_n\frac{\partial^2 Y_{lm}^*}{\partial\hat{L}_n\partial \hat{\mathbf{L}}}\times \hat{\mathbf{L}}\right)} < \frac{\xi_{\rm d}}{a_{\rm d}}.
%\end{equation}
\begin{equation}\label{eqn:stability criterion completely general}
    \abs{\sum_{lm} \sum_{n\in\set{x,y,z}}\frac{J_{lm}}{GM_{\rm d}}\Delta L_n\left(\frac{\partial^2 Y_{lm}^*}{\partial\hat{L}_n\partial \hat{\mathbf{L}}}\times \hat{\mathbf{L}}\right)\cdot\Delta \hat{\mathbf{L}}} < \frac{\xi_{\rm d}}{a_{\rm d}}\;,
\end{equation}
%where we take the norm of the vector whose $i$-th component is described here, $\eps_{ij}^{~~k}$ is the Levi-Civita symbol, and a sum over $j,k,n$ is implied. 
where $\Delta L_n$ is the $n$th Cartesian component of $\Delta \mathbf{L}$.
This holds for any $\Delta \mathbf{L} \perp \mathbf{L}_{\rm d}$, whose magnitude is $\Delta L = 2L\sin \iota $. For the case of equation \eqref{eqn:potential eigenvector}, this reduces to}
%\begin{equation}\label{eqn:stability general wave-function}
%    \abs{\sum_{l,l_0} \frac{2J_{l,l_0}}{GM_{\rm d}}\left[P_l''\left[\hat{\mathbf{L}}\cdot \hat{\mathbf{L}}_{\rm a}\right]\sin^2\theta - P_l'\left[\hat{\mathbf{L}}\cdot \hat{\mathbf{L}}_{\rm a}\right] \cos\theta\right]} < \frac{\xi_{\rm d}}{a_{\rm d}},
%\end{equation} 
\begin{equation}\label{eqn:stability general wave-function}
    \abs{\sum_{l,l_0} \frac{2J_{l,l_0}}{GM_{\rm d}}P_l''\left(\hat{\mathbf{L}}\cdot \hat{\mathbf{L}}_{\rm a}\right)\sin^2\theta} < \frac{\xi_{\rm d}}{a_{\rm d}},
\end{equation} 
where \chg{$J_{l, l_0}$ is defined below equation~\eqref{eqn:potential eigenvector}}, $\cos\theta = \hat{\mathbf{L}}_{\rm d}\cdot \hat{\mathbf{L}}_{\rm a}$\chg{, and the direction of $\Delta \mathbf{L}$ which maximises the left-hand side is in the plane spanned by $\mathbf{L}_{\rm d}$ and $\mathbf{L}_{\rm a}$}. 
\chg{Inequalities (\ref{eqn:stability criterion completely general}--\ref{eqn:stability general wave-function}) are the main theoretical result of this paper, and describe the disc's stability under the influence of any FDM configuration, and will apply to any nuclear stellar discs, including in other galaxies.}

\paragraph{Rotating core.}
\chg{Let us now apply the above result to a specific model of the FDM core, which we motivate below.} 
For a given axion core, one can approximate the mass of the gravitational atom \cite{Huietal2019,Bamberetal2021} as the mass enclosed in a sphere whose radius is the minimum between $R_{\rm c}$ and the radius of influence of the SMBH, $GM_\bullet/\sigma^2$, where $\sigma \approx 100\; \textrm{km s}^{-1}$ is the velocity dispersion of the nuclear cluster \cite{Schoedeletal2009}, that is,
\begin{equation}\label{eqn: atom mass from core mass}
	M_{\rm a} \approx \chg{M_{\rm a, exp} \equiv } M_{\rm c}\left(\frac{\min\set{\chg{GM_\bullet/\sigma^2},R_{\rm c}}}{R_{\rm c}}\right)^3. 
\end{equation}

Let us proceed to estimate the gravitational atom's \chg{shape}, for $m_a \leq 10^{-18} \; \textrm{eV}$. 
Generically, the core of an FDM halo should retain some rotation, because the halo \chg{should have} some angular momentum, which can be encapsulated by the spin parameter $\lambda \equiv L_{\rm h}\abs{E}^{1/2}/\left(GM_{\rm vir}^{5/2}\right)$, where $E$ is the binding energy of the halo %\bk{could we define $E$, is this the energy of the dark matter halo? Could we say if $\lambda\leq 1$ must hold in general?}\byg{We could, using Sundman's inequality and the virial theorem. I'll check the quantum-mechanical version.}
\citep[e.g.][]{Peebles1969,Bullocketal2001}.\footnote{One may bound $\lambda$ from above by a quantum-mechanical (loose) version of Sundman's inequality, $\abs{\sum_{ijk}\eps_{ijk}\bpsi x^jp^k\kpsi}^2 \leq 6\bpsi x^2\kpsi \bpsi p^2\kpsi$ \chg{combined with} the virial theorem.} With it, one can relate the global FDM halo properties to the rotation of the axion core (and hence the atom): %assuming the gravitational potential to be spherically symmetric, and because all axions share a single wave-function, 
one can \chg{define the parameter
\begin{equation}\label{eqn:alpha definition}
  \alpha \equiv \norm{\Pi_{l\neq 0} \kpsi} \sqrt{\frac{M_{\rm a}}{M_{\rm a, exp}}},
\end{equation}
where $\Pi_{l\neq 0}$ is an orthogonal projection on the $l\neq 0$ sub-space, $M_{\rm a}$ is the atom's mass, and $M_{\rm a, exp}$ is the value expected for it, i.e.~the one obtained from equations (\ref{eqn:core radius}--\ref{eqn: core mass}) and \eqref{eqn: atom mass from core mass}. $\alpha$ thus quantifies the `rotating mass' of the FDM atom, given by $\abs{\alpha}^2 M_{\rm a, exp}$. 
It is, strictly speaking, a free parameter of the model (physically determined by the intricate details of the atom's formation process).}

\chg{We now give an argument for a plausible value for $\alpha$, but later we treat it as an additional parameter and infer constraints on the pair $(m_a,\alpha)$. If $M_{\rm a} = M_{\rm a, exp}$,} the gravitational atom's wave-function \eqref{eqn:wave-function} satisfies
\begin{equation}
  \abs{\alpha}^2 \chg{=} \sum_{n,l\neq 0,m} \abs{\alpha_{nlm}}^2.
\end{equation}
%\chg{If} $\lambda \ll 1$ (and hence $\abs{\alpha} \ll 1$), for otherwise the gravitational potential would not be spherically symmetric, even approximately. 
For the core, one would have $\expec{L_{\rm a
}} \equiv \chg{M_{\rm c}/m_a}\bpsi L_{\rm a} \kpsi = GM_{\rm c}^{5/2}\lambda_{\rm c}/\sqrt{\abs{E_{\rm c}}}$, 
where $\lambda_{\rm c}$ denotes the core's spin parameter; this allows one to gauge $\alpha$ as follows. We assume that, energetically, the Bose-Einstein condensate would settle onto the lowest-energy states, subject to the constraint of fixed angular momentum $\expec{L_{\rm a
}} = GM_{\rm c}^{5/2}\lambda_{\rm c}/\sqrt{\abs{E_{\rm c}}}$. \chg{We argue in the appendix why, based on energy arguments,}
%If we orient the $\zhat$ axis to align with the direction of $\expec{\mathbf{L}_{\rm a}}$, then by the variational principle for the wave-function \cite[][\S 20]{LandauLifshitzQM}, one would have $\expec{nlm|\psi} =0$ for $m\neq 1, n\geq 2$, because that will allow the largest overlap $\abs{\expec{100|\psi}}$ with the (spherically symmetric) ground state, and thus the lowest total energy.\footnote{More accurately, $\sum_{n\geq 2, l\neq 0,m}\abs{\expec{nlm|\psi}}^2 \ll \abs{\expec{211|\psi}}^2$.} %\bk{the previous statement is unclear. Could we justify better or state that this is an intuitive expectation?}\byg{Is the way it's phrased now better?}\bk{If you think this will convince the referee then fine, otherwise more details could go into a footnote. It is unclear why strictly zero and not just smaller than other terms? For instance, an infinitely thin tilted disk would have all other terms or would it not? If it is smaller, is their sum also smaller? Finally, you mean the absolute value, right?} 
it is plausible to take
\begin{equation}\label{eqn:wave function typical}
	\kpsi \approx \alpha_{100}\ket{100} + \alpha_{211}\ket{211}
\end{equation}
for the \chg{field configuration} of the gravitational atom, where $\abs{\alpha_{100}}$, and $\abs{\alpha_{211}}$ are fixed by the above requirement on angular momentum. %\footnote{Other terms with different $m$ for the same $l=1$ are set to $0$ due to translation invariance.} 
\chg{While the state \eqref{eqn:wave function typical} is simple, the formalism developed here allows one to study the interaction of the atom with the disc for any $\psi$.}
%Correspondingly, the `rotating mass' is $M_{\rm a}\abs{\alpha}^2$. 
\chg{Thus,} $\lambda_{\rm c}$ \chg{determines} $\alpha$ by
\begin{equation}\label{eqn: rotating mass from core properties}
  \abs{\alpha}^2 = \lambda_{\rm c} \sqrt{\frac{10}{3}}\frac{m_a\sqrt{GM_{\rm c}R_{\rm c}}}{\sqrt{l_0(l_0+1)}\hbar} \approx 8.4\times 10^{-2} \left(\frac{\lambda_{\rm c}}{0.06}\right);
\end{equation}
here, the $m_a$-dependence drops, and we have used $\abs{E} = \frac{3GM_{\rm c}^2}{10R_{\rm c}}$ for a uniform, virialised core, and $L_{\rm a} = \hbar \abs{\alpha}^2\sqrt{l_0(l_0+1)}M_{\rm c}/m_a$ with $l_0=1$.\footnote{\chg{This could be relaxed to $L_{\rm a} = \hbar \abs{\alpha}^2m_0M_{\rm c}/m_a$, with $m_0 = 1$, if the angular-momentum vector is aligned with the $\zhat$-axis, but to be more conservative we keep the additional $\sqrt{2}$ factor.}} \chg{For our model of $\alpha$} we assume that \chg{it} is given by equation \eqref{eqn: rotating mass from core properties} with $\kpsi$ given by \eqref{eqn:wave function typical}, with the additional input that $\lambda_{\rm c} = \lambda_{\rm halo}$ \cite[cf.][]{Schobesbergeretal2021}. \chg{F}or the Milky Way, $\lambda_{\rm halo}$ is estimated to be $\lambda_{\rm halo} \approx 0.05$--$0.09$ \cite{Obrejaetal2022}, so here we use $\lambda_{\rm c} \approx 0.05$--$0.09$.\footnote{Ref.~\cite{Obrejaetal2022} estimated $\lambda' \equiv L_{\rm h}/[\sqrt{2}M_{\rm vir}R_{\rm h}V_{\rm circ}(R_{\rm h})]$, rather than $\lambda$, to be $0.061_{-0.016}^{+0.022}$. $\lambda'$ is smaller than $\lambda_{\rm halo}$ by a function of the concentration \cite{Moetal1998,Bullocketal2001}; for a halo concentration of $10$ \cite{Cautunetal2020} (before disc formation and halo contraction), this implies that $\lambda = 1.12\lambda'$, which is what we use.} The angle between the angular-momentum vector of the nuclear inner disc and the rotation direction of the Galactic disc, which we assume to be aligned with the halo and hence with $\hat{\mathbf{L}}_{\rm a}$, is approximately $\theta \approx 79^\circ$ \cite{Bartkoetal2009,Fellenbergetal2022}.

%Equations \eqref{eqn: atom mass from core mass} and \eqref{eqn: rotating mass from core properties} implies that for $m_a = 10^{-20} \; \textrm{eV}$, the rotating mass is $\abs{\alpha}^2 M_{\rm a}\approx 5000\; M_\odot$. 

The state $\ket{211}$ is also unstable to the super-radiant instability \cite{Zeldovich1971,PressTeukolsky1972,Cardosoetal2004}, and also potentially susceptible to accretion into the SMBH. We show in the appendix that neither of these phenomena is relevant for masses considered \chg{here}. 

\chg{In summary, we assumed that %the gravitational atom is such that 
(i) $\kpsi$ is well-approximated by equation \eqref{eqn:wave function typical} (justified by energy considerations), (ii) $M_{\rm a} = M_{\rm a,exp}$, (iii) $\lambda_{\rm c} = \lambda$, and (iv) the value of $\lambda$. By making $\alpha$ a free parameter, it absorbs the modelling assumptions (ii--iv) on its mass and angular momentum (and consequently the relation between $\lambda$ and $\lambda_{\rm c}$), which are thus relaxed.}

\begin{figure}
	\centering
	\includegraphics[width=0.49\textwidth]{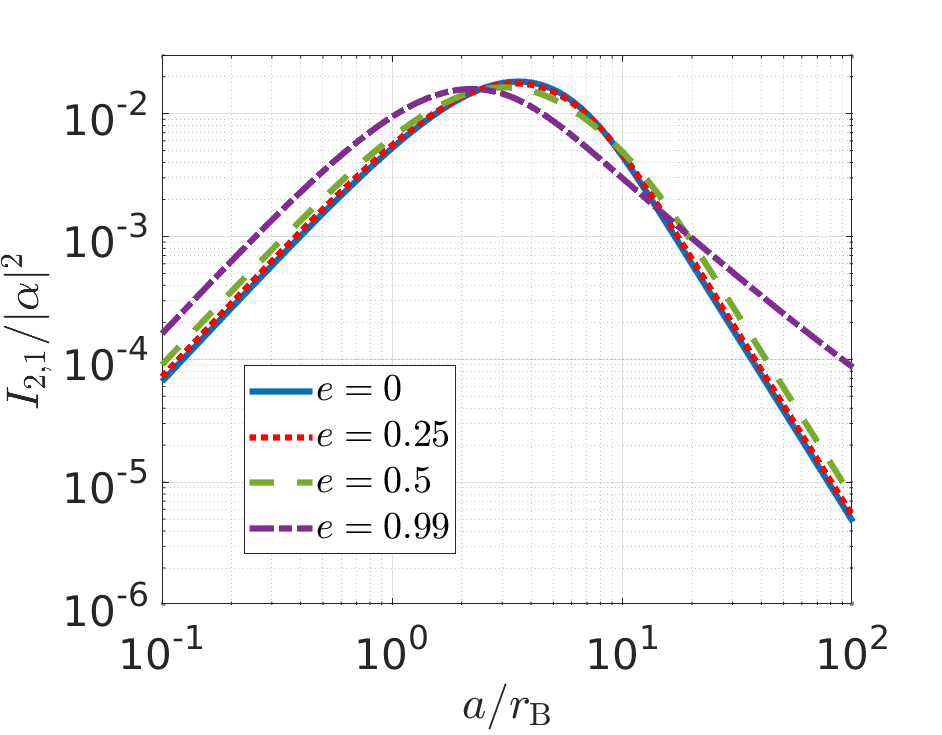}
	\caption{The integral $I_{2,1}$ as a function of the semi-major axis, plotted for various eccentricities, for the wave-function \eqref{eqn:wave function typical}.}
	\label{fig:I_21}
\end{figure}

For the state \eqref{eqn:wave function typical} equation \eqref{eqn:potential eigenvector} becomes
\begin{equation}\label{eqn:potental VRR 211}
    \expec{\Phi}_{\rm da} = \chg{\sum_{l_0=0}^1}\sum_{l=0}^{2\chg{l_0}}J_{l,\chg{l_0}}P_l\left(\hat{\mathbf{L}}\cdot \hat{\mathbf{L}}_{\rm a}\right),
\end{equation}
and by equations \eqref{eqn: I definition from J} and \eqref{eqn:J lm},
\begin{align}
	I_{0,0} & = 4\sqrt{2}\left[1-\abs{\alpha}^2\right]\int_0^\infty \frac{y^2 \mathrm{e}^{-2y/r_{\rm B}}s_{0}(a,y,e,0)}{r_{\rm B}^2\max\set{a,y}}\mathrm{d}y,   \\ 
	I_{2,1} & = \frac{\abs{\alpha}^2}{240r_{\rm B}^4} \int_0^\infty \frac{y^{4} s_{2}(a,y,e,0)}{\mathrm{e}^{y/r_{\rm B}}\max\set{a,y}} \left[\frac{\min\set{a,y}}{\max\set{a,y}}\right]^{2}\mathrm{d}y,
\end{align}
and the rest of $J_{l,l_0}$  are zero or negligible. 
$I_{2,1}$ is plotted in Figure \ref{fig:I_21}. 
The monopole term, $J_{0,0}$, sources mass precession \chg{of the pericentre} by the density profile $\rho_{\rm a} = M_{\rm a}\abs{\psi}^2$, while the quadrupole, $J_{2,1}$ is a VRR interaction. %Besides, $J_{1,1} = 0$ as expected from the equivalence principle. 

The potential \eqref{eqn:potental VRR 211} drives the precession of $\hat{\mathbf{L}}$ about the $\zhat$-axis, i.e. about $\hat{\mathbf{L}}_{\rm a}$, with frequency \cite{KocsisTremaine2015} 
\begin{equation}
	\boldsymbol{\Omega} = -3m_s\frac{J_{2,1}}{L}\left(\hat{\mathbf{L}}\cdot \hat{\mathbf{L}}_{\rm a}\right)\hat{\mathbf{L}}_{\rm a}. 
\end{equation}
%The mass-precession rate is 
%\begin{equation}\label{eqn: mass precession rate}
%	\omega_{\rm mp} = -\frac{GM_{\rm a}m_s}{r_{\rm B}}\frac{\sqrt{1-e^2}}{eL}\left(\frac{\mathrm{d}I_{0,0}}{\mathrm{d}e} + \frac{\mathrm{d}I_{0,1}}{\mathrm{d}e}\right),
%\end{equation}
%where we have neglected the quadrupole contribution, which is suppressed relative to the monopole. 

%\byg{Add figure that plots $\left[J_{0,0}r_{\rm B}/GM_{\rm a}\right]$ and $\left[J_{2,1}r_{\rm B}/GM_{\rm a}\lambda\right]$ for various eccentricities, as a function of the ratio $a/r_{\rm B}$.}

\paragraph{\chg{Constraints from disc stability.}}
For the case of the wave-function \eqref{eqn:wave function typical}, \chg{equation \eqref{eqn:stability general wave-function}} becomes 
\begin{equation}\label{eqn:stability}
	\abs{\alpha}^2M_{\rm a\chg{,exp}} < M_{\rm d}\frac{r_{\rm B}\xi_{\rm d}\abs{\alpha}^2}{a_{\rm d} 6I_{2,1}\abs{\chg{\sin^2}\theta}}.
\end{equation}
\chg{If the disc is unstable, we estimate its disruption time to be $\Omega_{\rm d}^{-1}$ \cite{PanamarevKocsis2024}, which is less than $10\; \textrm{Myr}$ for $m_a \in [3,7]\times10^{-20}\; \textrm{eV}$ (for the mean values of the parameters). } 

\begin{figure}
    \centering
    \includegraphics[width=0.49\textwidth]{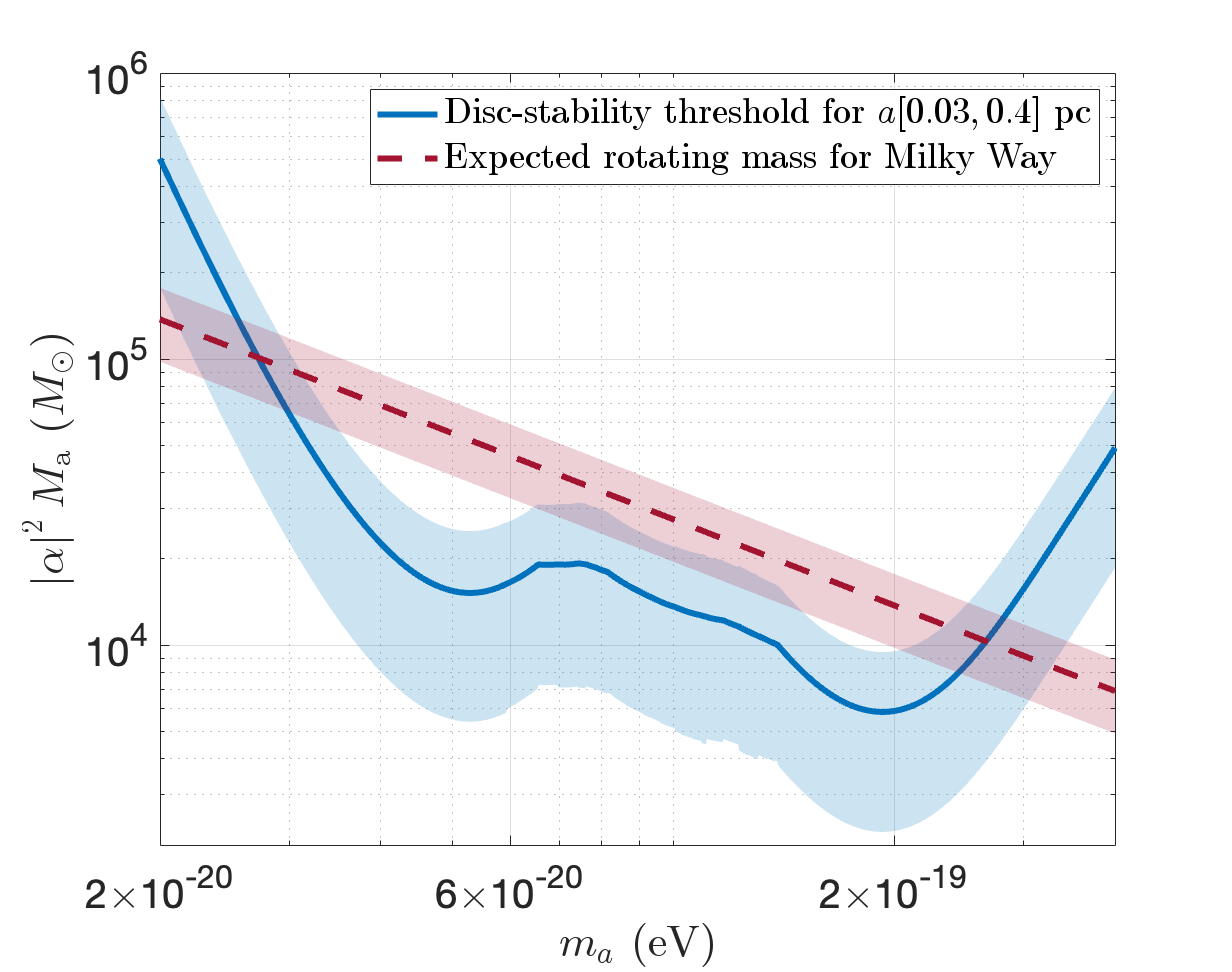}
    \caption{The upper limit for disc stability from inequality \eqref{eqn:stability}, for a Galactic-centre-like disc with parameters specified in the text and table \ref{tab:Parameters} (in blue). The expected value of the rotating mass $\abs{\alpha}^2M_{\rm a}$ is plotted too (red, dashed line). The shaded regions around both curves show estimated $1\sigma$ uncertainties. The disc should be disrupted for values of $m_{a}$ where the maximum rotating mass for disc stability is lower than $\abs{\alpha}^2M_{\rm a}$, and are thus constrained. %The black curves are the same contours of disc stability, but for $\iota = 15^\circ$, for comparison. \byg{Given that $15^\circ$ is less favoured by the data, should we show the black lines here, or do they obscure our point?}\bk{Perhaps the blue line is sufficient, we could put in words that the blue region shifts by a factor of XX upwards if $\iota = 15^\circ$ instead of $20^\circ$. BTW, what is the de-Broglie wavelength for these masses? Is it meaningful to put this on the top axis? Is there a minimum mass for the dark matter to be fuzzy?}\byg{'Fuzziness' is a fuzzy term, so no. I will add $r_{\rm B}$ on the top axis.}
    }
    \label{fig:constraint fluff}
\end{figure}

\begin{figure}
    \centering
    \includegraphics[width=0.49\textwidth]{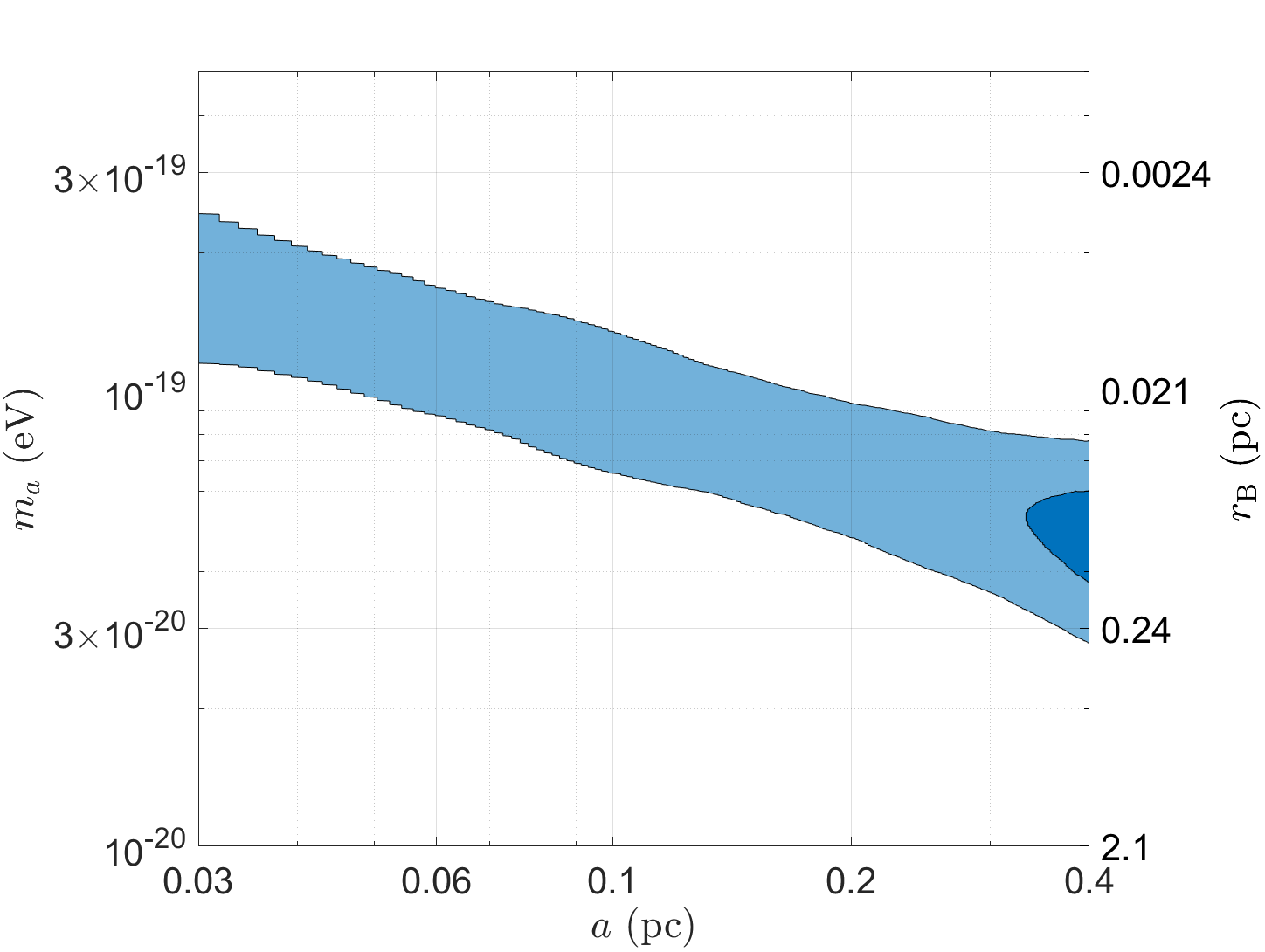}
    \caption{The semi-major axes in the disc which the gravitational atom renders unstable. Here $M_{\rm d} = 4000M_\odot$, $\iota = 10^\circ$, $M_{\rm vir} = 1.1\times 10^{12}\, M_\odot$, and $\lambda_{\rm c} = 0.06$; the inner, darker region corresponds to $M_{\rm d} = 10^4M_\odot$ with the same values for the other parameters.}
    \label{fig:breaking region}
\end{figure}

Th\chg{e} constraint \chg{\eqref{eqn:stability}} is plotted in figure \ref{fig:constraint fluff}; parameter values are specified in table \ref{tab:Parameters}. Here, we have used a disc whose semi-major-axis distribution is $\mathrm{d}N/\mathrm{d}a \propto a^{-3/4}$, the eccentricities are uniformly distributed, and the inclinations are normally-distributed with a variance $\iota$. We consider two possibilities for its value: $\iota = 14^\circ \pm 4^\circ$ \cite{PaumardEtAl2006} (a conservative value) and $\iota = 16^\circ$, preferred by recent observations \cite{Bartkoetal2009,Fellenbergetal2022},\footnote{\chg{Even if the initial inclination is smaller, it increases rapidly (within $\mathcal{O}(1)$ Myr), mostly by two-body relaxation within the disc, which is efficient in two dimensions \cite{SubrHaas2014,Panamarev2022}; external effects like the nuclear cluster would make this even more rapid \cite{Panamarev2022}.}} where we estimate the error as $4.8^\circ$;\footnote{This is done by considering the fourth moment of the haversine distance (on the angular-momentum-direction sphere) between the $15$ stars with known orbital parameters listed in \cite{Fellenbergetal2022}, and the location of the disc's centre at $(i,\Omega) = (126^\circ,104^\circ)$ \cite{Bartkoetal2009}.} %\byg{I am using 20 degrees, because this is what Sebastiano has in \cite{Fellenbergetal2022}, but this is really the key parameter that matters for everything.}\bk{Should we have another plot in the appendix for different values of $\iota$?}\byg{I could, e.g. for $\iota = 15^\circ$, but I think the observations favour $20^\circ$, don't they? I think this is a lower limit on $\iota$, which is twice the standard deviation of the angular-momentum vectors of the (inner) clockwise disc stars \cite{PaumardEtAl2006,Bartkoetal2009}. The former states that the standard deviation is $14^\circ \pm 4^\circ$, i.e. $\iota = 28^\circ \pm 8^\circ$, while \cite{Bartkoetal2009} say that the standard deviation is $16^\circ$---even larger.}\bk{Agreed, the current numbers are fine.}
%\footnote{This is a lower limit on $\iota$, which is twice the standard deviation of the angular-momentum vectors of the (inner) clockwise disc stars \cite{PaumardEtAl2006,Bartkoetal2009}. %Ref.~\cite{Fellenbergetal2022} lists 15 stars with measured orbital parameters in the clockwise disc. We computed the covariance matrix $\Sigma_{i,\Omega}$ of their inclinations and arguments of ascending node; the value of $\iota$ we use here is twice the mean of the square roots of the two eigenvalues of $\Sigma_{i,\Omega}$, which is $20.9^\circ \pm 5.4^\circ$ (assuming Poisson errors).
%} 
%for this profile $\xi_{\rm d} =1.02$.
The test star $s$ is placed at semi-major axes $a \in [0.03,0.4]\,\textrm{pc}$, with $\hat{\mathbf{L}}\cdot\hat{\mathbf{L}}_{\rm d} = \cos \iota$. \chg{Here, $\xi_{\rm d}$ is calculated for each value of $a$ directly from equation \eqref{eqn:disc precession frequency}, and for concreteness we used $648$ particles in the disc, normalised to have a total disc mass $M_{\rm d} = \sum_{i=1}^{648} m_i$, with equal masses.}\footnote{\chg{We truncated the multipole sum at $l=120$, for accuracy \cite{KocsisTremaine2015}.}}
The red line in figure \ref{fig:constraint fluff} is the expected value of $\abs{\alpha}^2M_{\rm a}$ based on the parameters described above, while the blue line corresponds to the value of $\abs{\alpha}^2M_{\rm a}$ saturating inequality \eqref{eqn:stability}, minimised over all semi-major axes of the test star. 
Uncertainties resulting from the errors in the parameters in table \ref{tab:Parameters} are plotted as shaded regions. An axion core with a particle mass $m_a$ such that the former curve lies above the latter is inconsistent with the existence of the clockwise disc at the Galactic centre, because such a disc would have been broken by the influence of the FDM atom. Figure \ref{fig:breaking region} shows which semi-major axes in the disc are expected to become unstable, in an example. 
Propagating the uncertainties yields that this occurs at $2\sigma$ for $m_a \in [m_{\rm low},m_{\rm high}]$, where $\set{m_{\rm low},m_{\rm high}} = \set{4.\chg{2},5.\chg{48}}\times 10^{-20}\,\textrm{eV}$ for the conservative choice of $\iota$, or $\set{\chg{3.82}, 5.\chg{9}8}\times 10^{-20}\,\textrm{eV}$ for $\iota = 16^\circ \pm 4.8^\circ$. 
%Figure \ref{fig:constraint fluff} shows that for the parameters described above, axion particle-masses in the range $m_{\rm low} \leq m_a \leq m_{\rm high}$, where $m_{\rm low} = 2.8\times 10^{-20}\,\textrm{eV}$ and $m_{\rm high} = 2.5\times 10^{-19}\,\textrm{eV}$, are inconsistent with the existence of the inner stellar disc in the Galactic centre, because such a disc would have been broken by the influence of the FDM atom. We have also attempted to address some of the modelling uncertainties, by considering disc masses between $3000\,M_\odot$ and $10^4\,M_\odot$, and $\lambda_{\rm c}$ between $0.05$ and $0.09$. These lead to constraints between $m_{\rm low} \in [2.14,4.1]\times 10^{-20}\,\textrm{eV}$ and $m_{\rm high} \in [5.6,32]\times 10^{-20}\,\textrm{eV}$. %\footnote{Besides $M_{\rm d}$, the other disc parameter these constraints are most sensitive to is the opening angle $\iota$: decreasing $\iota$ to $15^\circ$ changes these constraints to $m_{\rm low} \in [2.1,4]\times 10^{-20}\,\textrm{eV}$ and $m_{\rm high} \in [5.8,9.7]\times 10^{-20}\,\textrm{eV}$.}

\begin{table}
        \begin{tabular}{|c|c|c|}
        \hline
            Parameter & Value & Ref. \\
            \hline
            $M_{\rm vir}$ & $(1.1\pm 0.1)\times 10^{12} \,M_\odot$ & \cite{HuntVasiliev2025} \\
            $M_{\rm d}$ & $(6500 \pm 3500)\, M_\odot$ & \cite{PaumardEtAl2006,Bartko2010,Yelda2014} \\
            $\iota$ (option 1) & $14^\circ \pm 4^\circ$ & \cite{PaumardEtAl2006} \\
            $\iota$ (option 2) & $16^\circ \pm 4.8^\circ$ & \cite{Bartkoetal2009,Fellenbergetal2022} \\
            $\lambda$ & $0.067 \pm 0.018$ & \cite{Obrejaetal2022} \\
            \hline
        \end{tabular}
    \caption{Adopted parameter values and $1\sigma$ uncertainties.}
    \label{tab:Parameters}
\end{table}

%\begin{figure}
%	\centering
%	\includegraphics[width=0.49\textwidth]{constraint_clean}
%	\caption{The upper limit for disc stability from inequality \eqref{eqn:stability}, for various disc radii, where we used $a = R_{\rm d}$.}
%	\label{fig:constraint}
%\end{figure}

While these constraints did not account for the influence of the spherical nuclear cluster (which has some net angular momentum \cite{Fellenbergetal2022}), this is not expected to modify them significantly: the nuclear cluster is not aligned with the clockwise disc \cite{Schoedeletal2009,Feldmeier2014}, so if $m \in [m_{\rm low},m_{\rm high}]$, the atom alone breaks the disc and the nuclear cluster could not stabilise the disc\chg{---only destabilise it further}. On the other hand, if $m_a \gg m_{\rm high}$---and hence $M_{\rm a\chg{,exp}} \ll M_{\rm d}$---then it is possible that $\hat{\mathbf{L}}_{\rm a}$ would be rotated to align itself with $\mathbf{L}_{\rm d}$, by resonant dynamical friction \cite{Szolgyenetal2021,Ginatetal2023}; this, again, does not invalidate the conclusion for $m \in [m_{\rm low},m_{\rm high}]$. A full, detailed treatment of these situations is deferred for future work. %\bk{The previous two statements may not be clear and could be used as a criticism for rejecting the paper. Could we rephrase to make it clearer in which case is the prediction robust and in which case may the obtained constraint be possibly violated? I think that as long as we can claim a constraint for a given set of approximations consistent with observations, it is fine as long as we list all caveats.}\byg{I re-phrased. Is this better?}\bk{Yes, it is much better, thanks.}

\begin{figure}
    \centering
    \includegraphics[width=0.48\textwidth]{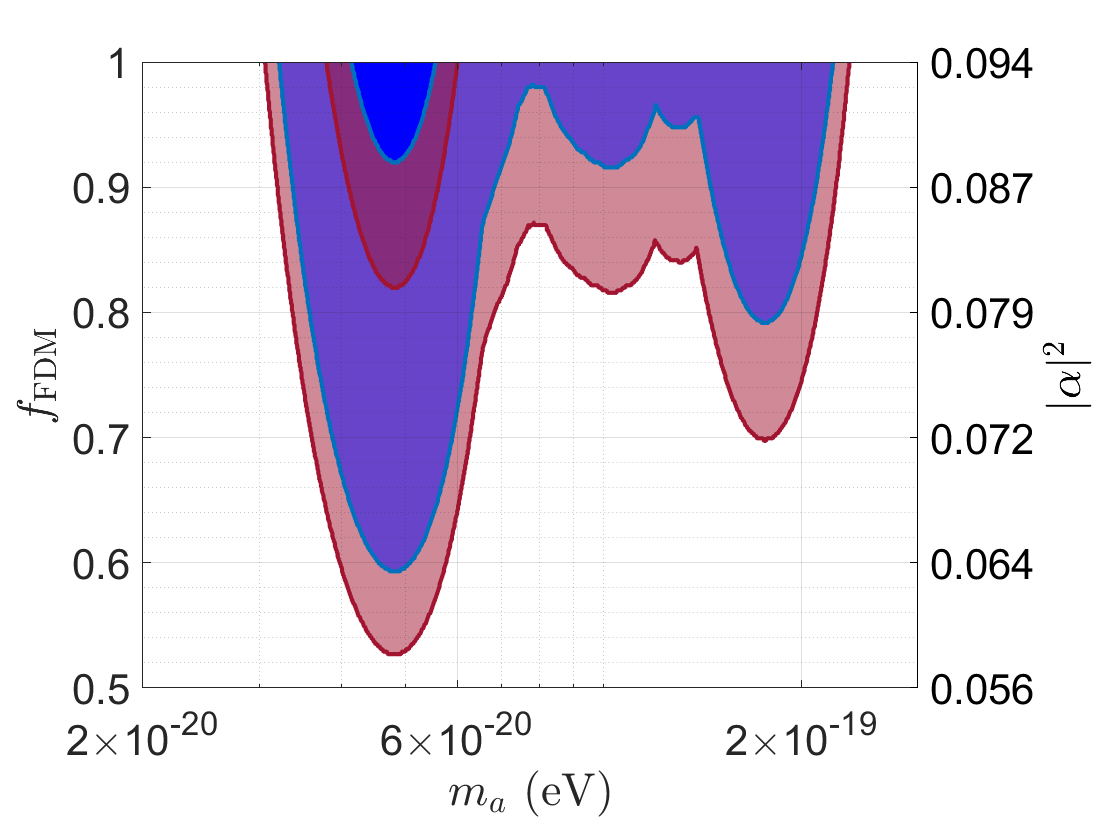}
    \caption{Values of $m_a$ \chg{inconsistent with a stable disc,} obtained from imposing inequality \eqref{eqn:stability}, for various values of \chg{$\alpha$ (or alternatively, fixing $\alpha$ by equation~\ref{eqn: rotating mass from core properties}, for} the FDM dark-matter fraction, $f_{\rm FDM}$\chg{)}. The shaded blue contours correspond to \chg{an opening angle} $\iota = 14^\circ \pm 4^\circ$, while the red ones use $\iota = 16^\circ \pm 4.8^\circ$. The darker regions are $2\sigma$ levels, while the lighter ones delineate $1\sigma$ regions\chg{, for the field configuration described in the text.}}
    \label{fig:constraint fraction}
\end{figure}

Generalising the dark-matter model to a case where only a fraction $f_{\rm FDM}$ of dark matter is fuzzy (and the rest is described by another model), one can generalise the derivation in this paper to derive a constraint on $m_{a}$ and $f_{\rm FDM}$. Assuming that the dark-matter core equations (\ref{eqn:core radius}--\ref{eqn: core mass}) continue to hold, $M_{\rm a}\propto f_{\rm FDM}^{4/3}$, while equation \eqref{eqn: rotating mass from core properties} and inequality \eqref{eqn:stability} are unchanged, we show in figure \ref{fig:constraint fraction} the corresponding constraint, generalising figure \ref{fig:constraint fluff}. %\bk{The previous sentence is unclear. Is this supposed to provide a more general extrapolation of the result for different assumptions? Which assumption is relaxed here? Could we explain more clearly what the point is?}\byg{I re-phrased. Is this better?}\bk{Yes. I tried to improve it, see I understood correctly.}
One can absorb $f_{\rm FDM}$ into $\alpha$, by transforming $\alpha \mapsto \alpha f_{\rm FDM}^{2/3}$, and then the contours in this figure can be construed as constraints on $\alpha$---because the right-hand side of \eqref{eqn:stability} is independent of both $\alpha$ and $f_{\rm FDM}$. This is therefore a way of incorporating uncertainties in equation \eqref{eqn: rotating mass from core properties}~\chg{and possible deviations from equations \eqref{eqn: core mass} and \eqref{eqn: atom mass from core mass}} into our analysis. \chg{By treating $\alpha$ as a free parameter, the stability of the disc only implies a joint constraint on the pair $(m_a,\alpha)$, via equation \eqref{eqn:stability}.}\footnote{\chg{Uncertainties in $\theta$ can also be absorbed into $\alpha$ by a re-definition} \chg{$\alpha \mapsto \alpha\sqrt{|\sin^2\theta /\sin^2 79^\circ|}$.}}
\chg{In this light, equation \eqref{eqn: rotating mass from core properties} can be viewed as a theoretical model for $\alpha$, which can break the degeneracy.}

\paragraph{Conclusions.}
In this letter, we showed that the dynamics of the Galactic centre are strongly influenced by ultra-light axions, which cannot be neglected in the dynamical modelling of the stellar distribution there---if these particles indeed constitute a sizeable fraction of dark matter. We calculated the gravitational torque they exert on stars, for a general axion-core state. We found that their influence is so strong, that under plausible assumptions on the rotation rate of the core and the gravitational atom's wave-function, these torques are should destabilise---or possibly break---the disc of young, massive stars. Given the observed properties of these stars in the Galactic centre, this imposes a stringent constraint on the allowed range of particle masses for the axions comprising this core\chg{, for the model studied here. Treating $\alpha$ as a free parameter to be constrained, allows one to relax many of the modelling assumptions, and incorporate them into the parameter $\alpha$, thereby producing $2$-dimensional constraints on $(m_a,\alpha)$}. These constraints will improve with better Galactic centre data---especially when the uncertainties in the parameters in table \ref{tab:Parameters} are reduced---and also potentially with observations of nuclear discs in other galaxies.

\paragraph{Acknowledgements.}
We are grateful to Kfir Blum, Sebastiano von Fellenberg, Pedro Ferreira, Sofia Flores, Chris Hamilton, John Magorrian and Taras Panamarev for helpful discussions. This work was supported by the Science and Technology Facilities Council grant No.~ST/W000903/1, and by a Leverhulme Trust International Professorship Grant (No.~LIP-2020-014). Y.B.G.~was partly supported by the Simons Foundation via a Simons Investigator Award to A.A.~Schekochihin.

%\software{astropy \citep{2013A&A...558A..33A,2018AJ....156..123A},          Cloudy \citep{2013RMxAA..49..137F}, Source Extractor \citep{1996A&AS..117..393B}}

\bibliography{axions,friction_bib}

\onecolumngrid
\appendix

\section{General \chg{FDM} Field}
\label{appendix: potential for general wave-function}
\paragraph{Potential.}
In this appendix we derive equation \eqref{eqn:axion potential VRR}, for a general Bose--Einstein condensate gravitational atom. 
Let the axion \chg{field} be
\begin{equation}
	\psi(\mathbf{y}) = \sum_{n,l,m} \alpha_{nlm}R_{nl}(y) Y_{lm}(\yhat),
\end{equation}
where $R_{nl}$ are the standard, hydrogen-atom radial wave-functions \cite[][\S 36]{LandauLifshitzQM},
whence the density is %\bk{shouldn't the left hand side also have a $y$ argument?}\byg{Of course. Fixed.}
\begin{equation}
	\rho(\mathbf{y}) = M_{\rm a}\sum_{\substack{n_1,l_1,m_1, \\ n_2,l_2,m_2}} \alpha_{n_1l_1m_1}\alpha^*_{n_2,l_2,m_2} R_{n_1l_1}(y)R_{n_2l_2}^*(y)Y_{l_1m_1}(\yhat)Y_{l_2m_2}^*(\yhat).
\end{equation}
%\bk{Is there a converse relation? Given $\rho$ what is $R_{n l}$?}\byg{$R_{nl}$ are the standard hydrogen radial wave-functions (it now says so above) -- a combination of an exponential and a Laguerre polynomial, so they are tabulated functions, but given $\psi(\mathbf{y})$ we can find $\alpha_{nlm}$ because the $R_{nl}$'s are orthogonal.}
Inserting this density and the spherical-harmonic decomposition of $\abs{\mathbf{x}-\mathbf{y}}^{-1}$ in equation \eqref{eqn:spherical harmonics 1/r} into equation \eqref{eqn:potential definition} gives an integral over a product of three spherical harmonics of $\yhat$. Let us start, then, with this angular integral, over $\yhat$, for each value of the quantum numbers $n_1,l_1,m_1,n_2,l_2,m_2$; this integral is \cite{Edmonds1960}
\begin{align}
& \int \mathrm{d}^2\yhat ~Y_{lm}(\yhat)Y_{l_1m_1}(\yhat)Y_{l_2m_2}^*(\yhat) = (-1)^{m_2}\int \mathrm{d}^2\yhat ~Y_{lm}(\yhat)Y_{l_1m_1}(\yhat)Y_{l_2,-m_2}(\yhat) \\ & 
= (-1)^{m_2} \sqrt{\frac{(2l+1)(2l_1+1)(2l_2+1)}{4\pi}}\left(\begin{array}{ccc}
l & l_1 & l_2 \\
0 & 0 & 0
\end{array} \right) \left(\begin{array}{ccc}
l & l_1 & l_2 \\
m & m_1 & -m_2
\end{array} \right),
\end{align}
where $\left(\begin{array}{ccc}
l_3 & l_2 & l_1 \\
m_3 & m_2 & m_1
\end{array} \right)$
%\begin{equation}
%\left(\begin{array}{ccc}
%l_3 & l_2 & l_1 \\
%m_3 & m_2 & m_1
%\end{array} \right)
%\end{equation}
is the $3$-$j$ symbol (which vanishes for $m_1+m_2+m_3\neq 0$). 

Inserting this into the potential $\Phi(\mathbf{x})$ in equation \eqref{eqn:potential definition} gives 
\begin{equation}\label{eqn:axion potential not averaged}
\begin{aligned}
	\Phi(\mathbf{x}) & = GM_{\rm a}\sum_{\substack{l,m, \\ n_1,l_1,m_1, \\ n_2,l_2,m_2}} (-1)^{m_2+1} \sqrt{\frac{4\pi(2l_1+1)(2l_2+1)}{(2l+1)}}\left(\begin{array}{ccc}
	l & l_1 & l_2 \\
	0 & 0 & 0
	\end{array} \right) \left(\begin{array}{ccc}
	l & l_1 & l_2 \\
	m & m_1 & -m_2
	\end{array} \right) \\ &
	\times \alpha_{n_1l_1m_1}\alpha^*_{n_2l_2m_2}\int_0^\infty \mathrm{d}y~y^2R_{n_1l_1}(y)R_{n_2l_2}(y)Y_{lm}^*(\xhat)\frac{\min\set{x,y}^{l}}{\max\set{x,y}^{l+1}}.
\end{aligned}
\end{equation}
This expression, while analytical, is too complicated to be of practical use. Let us now double-average it over the mean anomaly and the argument of pericentre, where $\mathbf{x}$ \chg{performs} of a Keplerian orbit about the origin. 
The problem is that, as $\kpsi$ is written in a specific basis, which comes with a specific co-ordinate system---a choice of a $\zhat$ axis for the gravitational atom---we are not free to choose the direction of $\xhat$. But, if $R = R(i,\Omega,\omega)$ is a rotation matrix which transforms the orbit from the $\xhat$-$\yhat$ plane to the reference frame used here, then by virtue of the spherical harmonics being an irreducible representation of $SO(3)$, 
\begin{equation}
	Y_{lm}^*(\xhat) = \sum_{m'=-l}^{l}\left[D^{(l)}_{mm'}(R)\right]Y_{lm'}^*(\xhat'),
\end{equation}
where now $\xhat'$ is in the $\xhat$-$\yhat$ plane, and $D(R)$ is Wigner's D-matrix \cite{Edmonds1960}.  

Now, that $\xhat'$ is oriented properly, the only terms affected by orbit-averaging are $Y_{lm'}^*(\xhat')\min\set{x,y}^{l}/\max\set{x,y}^{l+1}$, because the D-matrix only depends on $i$, $\Omega$ and $\omega$, which are fixed along an orbit. Indeed, the average 
\begin{equation}
\left\langle \frac{\min\set{x,y}^{l}}{\max\set{x,y}^{l+1}} Y_{lm'}^*(\xhat') \right\rangle_{\rm da}
\end{equation}
is equivalent to the double-average of a VRR interaction where one of the stars (with position $\mathbf{y}$) has a circular orbit. This is therefore completely equivalent to a two-ring interaction, which, fortunately, was already treated by ref.~\cite{KocsisTremaine2015}. The result is  
\begin{equation}
\left\langle \frac{\min\set{x,y}^{l}}{\max\set{x,y}^{l+1}} Y_{lm'}(\xhat')^* D^{(l)}_{mm'}(R)\right\rangle_{\rm da} = \frac{\sqrt{2l+1}}{\sqrt{4\pi}\max\set{a,y}}P_{l}(0)\delta^{\rm K}(m',0)\left[\frac{\min\set{a,y}}{\max\set{a,y}}\right]^{l} s_{l}(a,y,e,0)D^{(l)}_{m0}(R),
\end{equation}
\chg{where} $\delta^{\rm K}$ is the Kronecker delta-function, and $s_{l}$ is defined in ref.~\cite{KocsisTremaine2015}, and its value depends on whether $y < a(1-e)$, $a(1-e)\leq y \leq a(1+e)$, or $y > a(1+e)$. 

Setting $m'=0$ because of the delta-function also simplifies the D-matrix greatly, since
\begin{equation}
	D^{(l)}_{m0}(R) = \sqrt{\frac{4\pi}{2l+1}}Y_{lm}^*(i,\Omega),
\end{equation}
whence equation \eqref{eqn:axion potential not averaged} yields equation \eqref{eqn:axion potential VRR}, \emph{viz.},
\begin{equation}
	\expec{\Phi}_{\rm da} =- \sum_{l,m}J_{lm}Y_{lm}^*(\chg{\hat{\mathbf{L}}}),
\end{equation} 
where
\begin{equation}\label{eqn:J lm}
\begin{aligned}
	J_{lm} & = GM_{\rm a}\sum_{\substack{n_1,l_1,m_1, \\ n_2,l_2,m_2}} (-1)^{m_2} \sqrt{\frac{4\pi(2l_1+1)(2l_2+1)}{(2l+1)}}P_{l}(0)\left(\begin{array}{ccc}
	l & l_1 & l_2 \\
	0 & 0 & 0
	\end{array} \right) \left(\begin{array}{ccc}
	l & l_1 & l_2 \\
	m & m_1 & -m_2
	\end{array} \right) \\ &
	\times  \alpha_{n_1l_1m_1}\alpha^*_{n_2l_2m_2}\int_0^\infty \mathrm{d}y~\frac{y^2R_{n_1l_1}(y)R_{n_2l_2}(y)}{\max\set{a,y}} \left[\frac{\min\set{a,y}}{\max\set{a,y}}\right]^{l} s_{l}(a,y,e,0).
\end{aligned}
\end{equation}
By the properties of the $3$-$j$ symbol, the only non-zero contributions to $J_{lm}$ are from $m_2=m+m_1$ and $\abs{l_1-l_2} \leq l \leq l_1+l_2$. For instance, this implies that if $\kpsi$ is an eigenstate $\ket{n_0,l_0,m_0}$, then $\expec{\Phi}_{\rm da}$ only has terms with $m=0$ and $l\in \set{0,\ldots,2l_0}$. Then, equation \eqref{eqn:J lm} yields
\begin{equation}\label{eqn:J ll0}
	\frac{J_{l,l_0}}{GM_{\rm a}} = (-1)^{m_0} (2l_0+1)P_{l}(0)\left(\begin{array}{ccc}
	l & l_0 & l_0 \\
	0 & 0 & 0
	\end{array} \right) \left(\begin{array}{ccc}
	l & l_0 & l_0 \\
	0 & m_0 & -m_0
	\end{array} \right)   |\alpha_{n_0l_0m_0}|^2\int_0^\infty \mathrm{d}y~\frac{y^2R_{n_0l_0}^2(y)}{\max\set{a,y}} \left[\frac{\min\set{a,y}}{\max\set{a,y}}\right]^{l} s_{l}(a,y,e,0).
\end{equation}

We remark, that the energy levels of the gravitational atom are such that the frequencies are \citep{Baumannetal2019}
\begin{equation}
	\omega_{nlm} = \frac{E_{nlm}}{\hbar} \approx \frac{G^2 M_\bullet^2m_a^3}{2n^2\hbar^2} \gg {\rm Myr}^{-1}
\end{equation}
for order-unity $n$, so in-so-far as VRR is concerned, orbit-averaging should remove the quantum interference between different energy-levels, i.e. set $\alpha_{n_1l_1m_2}\alpha^*_{n_2l_2m_2} \mapsto 0$ in \eqref{eqn:J lm} if $n_1 \neq n_2$. 

\paragraph{Accretion and super-radiant instability.} 
If $m_{a} \gtrsim 10^{-18} \; \textrm{eV}$, then the time-scale for the super-radiant instability for the most unstable mode $\ket{211}$ is much shorter than a Hubble time, for $M_\bullet = 4\times 10^6\; M_\odot$, whence for $m_a \gtrsim 10^{-18} \; \textrm{eV}$, $\kpsi$ would be dominated by the fastest-growing super-radiant mode. %For Andromeda, $M_\bullet = 1.4\times 10^8 \; M_\odot$ \citep{Benderetal2005}, which means that the instability growth-rate would be smaller than a Hubble time for $m_a \gtrsim 4\times 10^{-20} \; \textrm{eV}$.
However, as seen in figures \ref{fig:constraint fluff} and \ref{fig:constraint fraction}, the constraints from the nuclear stellar disc are much stronger when $m_a < 10^{-18} \; \textrm{eV}$, even though in this case the instability's growth rate is too slow to matter, even over a Hubble time\chg{. We} thus keep $\alpha$ and $M_{\rm a}$ as given by equations \eqref{eqn: atom mass from core mass} and \eqref{eqn: rotating mass from core properties}, rather than $\alpha = 1$ (as would be the case for $m_a \gtrsim 10^{-18} \; \textrm{eV}$).

Note, that the $l=0$ modes of the soliton might be accreted by the SMBH, thereby weakening constraints from mass precession \cite{Huietal2019,Cloughetal2019,Baretal2019,DavisMocz2020,Bamberetal2021}: the decay rate due to accretion (conversely, the growth-rate of the super-radiant instability) of mode $\ket{nlm}$ is \cite{Baumannetal2019}
\begin{equation}
	\Gamma_{nlm} \sim \left(m \Omega_{\bullet} - \omega_{nlm}\right)\left(\frac{GM_\bullet m_a}{\hbar c}\right)^{4l+5},
\end{equation}
where $\omega_{nlm} = E_{nlm}/\hbar$ is the oscillation frequency of this mode, $\Omega_\bullet \equiv c^3\chi_\bullet\left[4GM_\bullet\left(1+\sqrt{1-\chi_\bullet^2}\right)\right]^{-1} \gg \omega_{nlm}$, and $\chi_\bullet$ is the dimension-less spin of the SMBH. This implies that, for $l=0$, $\Gamma_{n00}^{-1}$ is of the order of the age of the Universe (or smaller) for spherically symmetric modes, $M_\bullet = 4\times 10^6\,M_\odot$ and $m_a \gtrsim 5\times 10^{-20}\; \textrm{eV}$; but the $l\neq 0$ modes, with which we are concerned here, are protected by the extra power of $4l$, and hence their accretion time-scale is far longer than a Hubble time for $m_a < 10^{-18} \; \textrm{eV}$.%; similarly, in the unstable case, the super-radiant instability grows over a time-scale which is much longer than a Hubble time, as mentioned above. %\bk{What is $\Omega_{\bullet}$? How did we get the numerical value without specifying $\omega_{nlm}$ and $m\Omega_{\bullet}$? It is unclear if this expression is specifically for $l=0$ modes, as indicated by the start of the paragraph. Could we rephrase?}\byg{Yes, sorry. For $m=1$ this is dominated by $\Omega_\bullet = c^3/(4GM_\bullet)$. I'll rephrase.}

\paragraph{\chg{Wave-function shape.}}
\chg{Let us give an argument for equation \eqref{eqn:wave function typical}. Let us first consider the case where the soliton forms during halo formation, and the SMBH grows adiabatically as the Galaxy evolves. In that case, initially, well inside the core, the gravitational potential is that of a 3D harmonic oscillator, where the energy levels grow linearly with $n$. Thus, the total energy is roughly $E_0\left[1-\abs{\alpha}^2 + (n-1)\abs{\alpha}^2\right]$, where $E_0$ is the ground-state energy. If $\abs{\alpha}^2 \sim 1/l$ (cf.~equation \eqref{eqn: rotating mass from core properties}), with $l\leq n-1$, then the total energy is approximately independent of $l$. This neglects the self-gravity of the $l\neq0$ modes, which, when accounted for, would increase the energy by a term proportional to $\abs{\alpha}^4\left(1-\frac{\expec{r}_{n=1}}{\bpsi r \kpsi}\right)$, which is minimised by taking $\psi$ to include just the lowest possible states $n$ with $l\neq 0$, i.e.~only $n=1, 2$, with the magnitude of the $(n=2, l=1)$ component fixed by the angular-momentum constraint. This is without an SMBH, but by the adiabatic theorem \cite{Born:1928cqs}, this state would evolve into equation \eqref{eqn:wave function typical} as the SMBH grows. 
}

\chg{Let us now consider the opposite case, where the soliton forms with the SMBH \emph{in situ}. Let $\kpsi \propto \alpha_{n_1 00}\ket{n_1 00} + \sum_{l_0m_0}\alpha_{n_0l_0m_0}\ket{n_0l_0m_0}$, where $l_0 \neq 0$. The spherically symmetric $\ket{n_1 00}$ component will be accreted onto the SMBH, leaving only the states $\ket{n_0l_0m_0}$. Such a state has the lowest energy if $n_0$ is lowest, but for $l_0 \neq 0$, this requires $n_0 \geq 2$, with the minimum energy achieved by $n_0 = 2$, whence $l_0 = 1$. We assume that the system would relax to this state. If we orient the $\zhat$ axis to align with the direction of $\expec{\mathbf{L}_{\rm a}}$, then the conditions that $\expec{L_{{\rm a},x}} = \expec{L_{{\rm a},y}} = 0$ and $\expec{L_{{\rm a},z}} > 0$ fix $\alpha_{n_0l_00} = 0$. If $\expec{L_{{\rm a},z}}$ is fixed, then so is $\abs{\alpha_{211}}^2 - \abs{\alpha_{21-1}}^2$. Minimising $\abs{\alpha}^2 = \abs{\alpha_{211}}^2 + \abs{\alpha_{21-1}}^2$, subject to this constraint (to minimise the total energy), yields $\alpha_{21-1}=0$, $\alpha_{211} = \alpha$. Hence equation \eqref{eqn:wave function typical} arises in both cases.
}

\end{document}